\documentclass[twocolumn, times, showpacs,preprintnumbers,amsmath,amssymb,prl]{revtex4}

\usepackage{graphicx}%{epsfig}% Include figure files

\usepackage{dcolumn}% Align table columns on decimal point

\usepackage{bm}% bold math

\usepackage{times}% bold math

\begin{document}

\preprint{version 12b}
\title{New Limits on the Drift of Fundamental Constants from Laboratory
Measurements}
\author{M. Fischer}
\author{N. Kolachevsky }
\altaffiliation[Also at P.N. Lebedev Physics Institute, Moscow,
Russia] \
\author{M. Zimmermann}
\author{R. Holzwarth}
\author{Th. Udem}
\author{T.W. H\"{a}nsch }
\altaffiliation[Also at Ludwig-Maximilians-University, Munich,
Germany] \

\affiliation{Max-Planck-Institut f\"{u}r Quantenoptik (MPQ),
Hans-Kopfermann-Stra{\ss}e 1, 85748 Garching, Germany}

\author{M.~Abgrall$^1$}
\author{J.~Gr\"{u}nert$^1$}
\author{I.~Maksimovic$^1$}
\author{S.~Bize$^1$}
\author{H.~Marion$^1$}
\author{F.~Pereira Dos Santos$^1$}
\author{P.~Lemonde$^1$}
\author{G.~Santarelli$^1$}
\author{P.~Laurent$^1$}
\author{A.~Clairon$^1$}
\author{C.~Salomon$^2$}

\affiliation{**$^{1}$BNM-SYRTE, Observatoire de Paris, 61 Avenue
de l'Observatoire, 75014 Paris, France\\}
\affiliation{**$^{2}$Laboratoire Kastler Brossel, ENS, 24 rue
Lhomond, 75005 Paris, France}
\date{\today}

\author{M.~Haas$^3$}
\author{U.D.~Jentschura$^3$}
\author{C.H.~Keitel$^3$}
\affiliation{**$^3$ Max-Planck-Institut f\"{u}r Kernphysik, Saupfercheckweg 1, 69117 Heidelberg, Germany}

\begin{abstract}

We have remeasured the absolute $1S$-$2S$ transition frequency
$\nu_{\rm {H}}$ in atomic hydrogen. A comparison with the result
of the previous measurement performed in 1999 sets a limit of
$(-29\pm 57)$~Hz for the drift of $\nu_{\rm {H}}$ with respect to
the ground state hyperfine splitting $\nu_{{\rm {Cs}}}$ in
$^{133}$Cs. Combining this result with the recently published
optical transition frequency in $^{199}$Hg$^+$ against $\nu_{\rm
{Cs}}$ and a microwave $^{87}$Rb and $^{133}$Cs clock comparison,
we deduce separate limits on $\dot{\alpha}/\alpha = (-0.9\pm
2.9)\times 10^{-15}$~yr$^{-1}$ and the fractional time variation
of the ratio of Rb and Cs nuclear magnetic moments $\mu_{\rm
{Rb}}/\mu_{\rm {Cs}}$ equal to $(-0.5 \pm 1.7)\times
10^{-15}$~yr$^{-1}$. The latter provides information on the
temporal behavior of the constant of strong interaction.

\pacs {06.30.Ft, 06.20.Jr, 32.30.Jc}
\end{abstract}

\maketitle

In the era of a rapid development of precision experimental
methods, the stability of fundamental constants becomes a question
of basic interest. Any drift of non-gravitational constants is
forbidden in all metric theories of gravity including general
relativity. The basis of these theories is Einstein's Equivalence
Principle (EEP) which states that weight is proportional to mass,
and that in any local freely falling reference frame, the result
of any non-gravitational experiment must be independent of time
and space. This hypothesis can be proven only experimentally as no
theory predicting the values of fundamental constants exists. In
contrast to metric theories, string theory models aiming to unify
quantum mechanics and gravitation allow for, or even predict,
violations of EEP. Limits on the variation of fundamental
constants might therefore provide important constraints on these
new theoretical models.

 A recent analysis of quasar absorption
spectra with redshifted UV transition lines indicates a variation
of the fine structure constant $\alpha=e^2/4\pi\varepsilon_0\hbar
c$ on the level of $\Delta \alpha/\alpha = (-0.54\, \pm
\,0.12)\times 10^{-5} $ for a redshift range $(0.2 < z < 3.7)$
\cite{Mur03}. On geological timescales, a limit for the drift of
$\alpha$ has been deduced from isotope abundance ratios in the
natural fission reactor of Oklo, Gabon, which operated about
$2$~Gyr ago. Modeling the processes which have changed the isotope
ratios of heavy elements gives a limit of $\Delta \alpha/\alpha =
(-0.36\, \pm\, 1.44)\times 10^{-8}$ \cite{Fuj00}. In these
measurements, the high sensitivity to the time variation of
$\alpha$ is achieved through very long observation times at
moderate resolution for $\Delta\alpha$. Therefore, they are
vulnerable to systematic effects \cite{Uzan}.

Laboratory experiments can reach a $10^{-15}$ accuracy within
years with better controlled systematics. This type of experiment
is typically based on repeated absolute frequency measurements,
i.e. comparison of a transition frequency with the reference
frequency of the ground state hyperfine transition in $^{133}$Cs.
For an optical transition, the theoretical expression for the
drift of its absolute frequency inevitably involves
$\mu_{\rm {Cs}}/\mu_B$ in addition to $\alpha$, where $\mu_{\rm
{Cs}}$ is the magnetic moment of the Cs nucleus and $\mu_B$ is the
Bohr magneton  \cite{Karsh}. The microwave Rb and Cs clock
comparison \cite{Mar03} even involves two nuclear moments. The magnitude of nuclear
moments results from both the electromagnetic and the strong
interaction.

Contributions from weak, electromagnetic, and strong interactions
can be disentangled by combining several frequency measurements
possessing a different sensitivity to the fundamental constants.
In this letter, we deduce separate stringent limits for the drifts
of the fine structure constant $\alpha$,  $\mu_{\rm {Cs}}/\mu_B$
and $\mu_{\rm {Rb}}/\mu_{\rm Cs}$ from combining the drifts of two
optical frequencies in hydrogen and in the mercury ion with
respect to the ground state hyperfine splitting in $^{133}$Cs and
the result of a microwave clock comparison \cite{Mar03}. Comparing
measurements performed at different places and at different times
we have to use the Lorentz and position invariance and we have to
assume that the constants change linearly and do not oscillate on
a year scale. With exception of this, our results are independent
of further assumptions about a particular drift model and possible
correlated drifts of different constants \cite{Cal02}.

The experiments on the drift of the $^{199}$Hg$^+$ $5d^{10}6s\
{^{2}}S_{1/2} (F=0) \rightarrow 5d{^{\,9}} 6s^2\,{^{2}}D_{5/2}
(F'=2)$ electric quadrupole (``clock'') transition frequency
$\nu_{\rm {Hg}}$ were performed by the group of J.~Bergquist at
NIST between July 2000 and December 2002 \cite{Biz03}. These
repeated measurements of $\nu_{\rm {Hg}}$ limit the fractional
time variation of the ratio $\nu_{\rm {Cs}}/\nu_{\rm {Hg}}$ to $
(0.2 \pm 7 )\times 10^{-15}$~yr$^{-1}$.

\begin{figure}[t]
\begin{center}
\includegraphics[width=7.5cm]{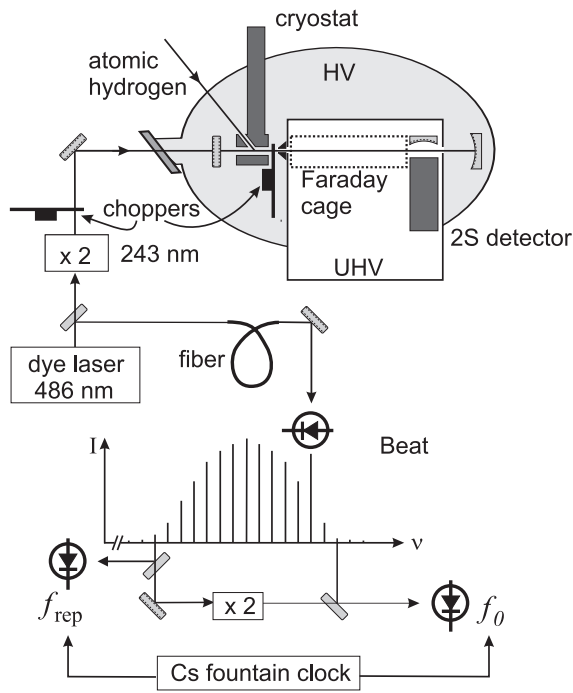}
\caption{\em Simplified experimental setup for comparison of the
hydrogen $1S$-$2S$ transition frequency with a primary frequency
standard (HV: high vacuum, UHV: ultra-high vacuum). A mode-locked
femtosecond laser with repetition rate $f_{\rm rep}$ emits the
frequency comb which is broadened in a photonic crystal fiber.
$f_0$ is the carrier-envelope offset frequency.} \label{hchain03}
\end{center}
\end{figure}

In 1999 \cite{Nie00a} and 2003, we have phase coherently compared
the frequency of the  $(1S, F\!\!=\!\!1, m_F\!\!=\!\!\pm 1)
\rightarrow (2S, F'\!\!=\!\!1, m_F'\!\!=\!\!\pm 1)$ two-photon
transition in atomic hydrogen to the frequency of the ground state
hyperfine splitting of $^{133}$Cs using a frequency comb technique
\cite{Rei00a}. The 1999 setup of the hydrogen spectrometer has
been described previously in \cite{Hub98b}, so we present only a
brief description with emphasis on the improvements since 1999. A
sketch of the actual setup is shown in Fig.\ref{hchain03}.

A cw dye laser emitting near 486 nm is locked to an external
reference cavity. The new cavity for the 2003 measurement is more
stable and drifts less than 0.5 Hz/s at 486 nm. The linewidth of
the dye laser stabilized on the new cavity has been characterized
as 60 Hz for an averaging time of 1~s. A small part of the laser
light is transferred to a neighboring lab where its absolute
frequency is measured. The main part is frequency doubled, and the
resulting radiation near 243 nm (corresponding to half of the
$1S$-$2S$ transition frequency) is coupled into a linear
enhancement cavity inside the vacuum chamber of the hydrogen
spectrometer.

Hydrogen atoms from a radio-frequency (rf) gas discharge are
cooled  to 5-6~K by collisions with the walls of a copper nozzle.
The nozzle forms a beam of cold atomic hydrogen which leaves the
nozzle collinearly with the cavity axis and enters the interaction
region between the nozzle and the detector shielded from stray
electric fields by a Faraday cage. Some of the atoms are
 excited from the ground state to the metastable $2S$
state by Doppler-free absorption of two counter-propagating
photons from the laser field in the enhancement cavity. The 1999
measurement was performed at a background gas pressure of around
$10^{-6}$ mbar in the interaction region. In the meantime, we
upgraded the vacuum system with a differential pumping
configuration. This allows us to vary the background gas pressure
in a range of $10^{-8}$-$10^{-7}$ mbar  and to reduce the
background gas pressure shift and the corresponding uncertainty to
2~Hz.

Due to small apertures, only atoms flying close to the cavity axis
can enter the detection region where the $2S$ atoms are quenched
in a small electric field and emit $L_{\alpha}$-photons. The
excitation light and the hydrogen beam are periodically blocked by
two mutually phase locked choppers, and the $L_{\alpha}$-photons
are counted time-resolved only in the dark part of a cycle. The
delay $\tau$ between blocking the 243 nm radiation and the start
of photon counting sets the upper limit for the atomic velocity of
$v < l/\tau$, where $l$ is the distance between nozzle and
detector. With the help of a multi-channel scaler, we count all
photons and sort them into 12 adjacent time bins. From each scan
of the laser frequency over the hydrogen $1S$-$2S$ resonance we
therefore get up to 12 spectra measured with different delays. To
correct for the second order Doppler shift, we use an elaborate
theoretical model to fit all the delayed spectra of one scan
simultaneously with a set of 7 fit parameters \cite{Hub98b}. The
result of the fitting procedure is the $1S$-$2S$ transition
frequency for a hydrogen atom at rest.

The transition frequency depends linearly on the excitation light
intensity due to the dynamic AC-Stark shift. We vary the intensity
and extrapolate the transition frequency to zero intensity to
correct for this effect \cite{Nie00a}.

For an absolute measurement of the $1S$-$2S$ transition frequency,
 the dye laser frequency near 616.5 THz is phase coherently
compared with an atomic cesium fountain clock. We take advantage
of the frequency comb technique recently developed in Garching and
Boulder \cite{Rei00a} to bridge the large gap between the optical
and radio frequency domains. A mode locked femtosecond (fs) laser
emits a pulse train which equals a comb of laser modes in the
frequency domain. The mode spacing corresponds to the repetition
rate frequency $f_{\rm rep}$ of the fs laser, and the frequency of
each mode can be written as $f_n = n f_{\rm rep}+f_0$, where $n$
is a large integer number and $f_0 < f_{\rm rep}$ is the comb
offset frequency. The repetition rate $f_{\rm rep}$ can easily be
detected with a photodiode. For combs spanning more than one
octave, $f_0$ can be determined by frequency doubling an infrared
mode $f_n$ in a nonlinear crystal to $2  f_n$ and measuring
the beat note with the appropriate mode $f_{2n}$ of the blue part
of the comb, yielding $f_0 = 2f_n-f_{2n}$.

\begin{figure}[b]
\begin{center}
\includegraphics[width=8cm]{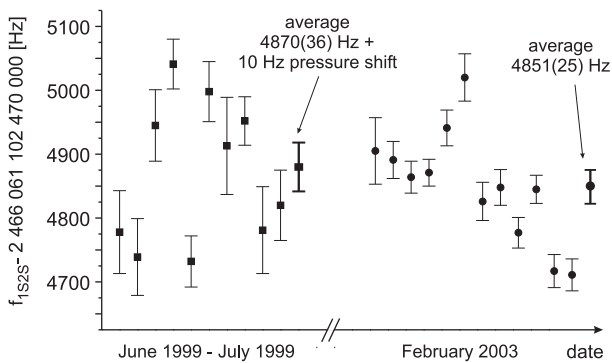}
\caption{\em Experimental results and averages for the 1999 and
2003 measurements of the absolute $(1S, F\!=\!1,
m_F=\pm1)\rightarrow(2S,F'\!=\!1, m'_F=\pm1)$ transition frequency
in atomic hydrogen. The background pressure shift of +10 Hz as
explained in \cite{Nie00a} is added to the 1999 value.}
\label{scatter9903}
\end{center}
\end{figure}

In our experiment, the frequency comb from a mode-locked fs
Ti:Sapphire laser with a repetition rate $f_{\rm rep }$ of 800 MHz
is broadened in a photonic crystal fiber to cover more than an
optical octave. We phase lock both degrees of freedom of the comb
($f_{\rm rep}$, $f_0$) to the radio frequency output of the Cs
fountain clock to get a frequency comb with optical modes whose
frequencies are known with the accuracy of the primary frequency
standard. Any of these modes is available for optical
measurements.

For both the 1999 and 2003 measurements, the transportable Cs
fountain clock FOM has been installed at MPQ. Its stability is
$1.8\times 10^{-13}\tau_{\rm ave}^{-1/2}$ for an averaging time $\tau_{\rm ave}$ and its accuracy has been
evaluated to $8\times 10^{-16}$ \cite{Abg03} at BNM-SYRTE. During
the  experiments in Garching, only a verification at the level of
$10^{-15}$ has been performed. Consequently we attribute a
conservative FOM accuracy of $2\times10^{-15}$ for these
measurements.

We have measured the $1S$-$2S$-transition in atomic hydrogen
during 10 days in 1999 and during 12 days in 2003. For
comparability, both data sets have been analyzed using the same
theoretical line-shape model~\cite{Hub98b}. In
Fig.\ref{scatter9903}, the results of the extrapolation to zero
excitation light intensity and the respective statistical error
bars for each day are presented. The statistical uncertainty was
significantly reduced compared to the 1999 measurements due to the
narrower laser linewidth and a better signal-to-noise ratio, but
the scatter of the day averages did not reduce accordingly. We
ascribe this effect to a residual uncompensated first-order
Doppler shift. Later measurements performed without the fountain
clock with a deliberately introduced asymmetry in the 243 nm
cavity indicate an adjustment-dependent frequency shift (the
elimination of this additional systematic shift should be of high
concern in future measurements). The effect should average out
after multiple re-adjustments of the spectrometer which have been
typically performed twice a day. It is impossible to correct the
data {\em a posteriori} because such details of the spectrometer
adjustment were not recorded during the phase-coherent
measurement. Other effects which can cause the systematic shift
(intra-beam pressure shift, background gas pressure shift, Stark
shift of the hyperfine levels induced by the rf gas discharge,
stray electric fields) have been checked and can be excluded on
the level of the observed scatter.

The 1999 and 2003 day-dependent data were averaged without
weighting \cite{remark}.  After accounting for the total
systematic uncertainties of 28~Hz (1999) and 23~Hz (2003) in the
mutually equivalent evaluation processes of  both measurements we
deduce a difference equal to $(-29\pm 57)$~Hz over a time interval
of 44 months. This is equivalent to a fractional time variation of
the ratio $\nu_{\rm Cs}/\nu_{\rm H}$ equal to $(3.2\pm6.3) \times
10^{-15}$ yr$^{-1}$.

The frequency of any optical transition can be written as
 $\nu=A\,  Ry \, F_{\rm rel}(\alpha)$ \cite{Dzu99,Karsh}, where
$Ry$ is the Rydberg energy and the relativistic correction $F_{\rm
rel}(\alpha)$ takes into account relativistic and many-body
effects. $F_{\rm rel}$ depends on the transition in the system
considered and embodies the dependence on $\alpha$, while the
parameter $A$ is independent of fundamental constants. For
absolute frequency measurements $Ry$ always cancels out.

Numerical calculations of the dependence of $F_{{\rm rel},
{\rm{Hg}}}(\alpha)$ for $\nu_{\rm {Hg}}$ on the fine structure
constant $\alpha$ yield \cite{Dzu99}
\begin{equation}
\alpha \frac{\partial}{\partial \alpha}\ln F_{{\rm rel},
{\rm{Hg}}}(\alpha)\approx\frac{\alpha}{\Delta \alpha}\frac{\Delta
F_{\rm rel}}{F_{\rm rel}} = -3.2\, .
\end{equation}

The corresponding expression for $\nu_{\rm {H}}$ gives a very weak
dependence on $\alpha$:
\begin{equation}
\alpha \frac{\partial}{\partial \alpha}\ln F_{{\rm rel},
{\rm{H}}}(\alpha) \approx 0\,.
\end{equation}
The frequency of the ground state hyperfine transition in
$^{133}$Cs is given by
\begin{equation} \nu_{\rm {Cs}} = A'\, Ry \,  \alpha^2\,
\frac{\mu_{\rm {Cs}}}{\mu_{B}} \, F_{{\rm rel},
{\rm{Cs}}}(\alpha).
\end{equation}
Following \cite{Dzu99}, the relativistic correction $F_{{\rm rel},
{\rm{Cs}}}(\alpha)$ is
\begin{equation}
\alpha \frac{\partial}{\partial \alpha}\ln F_{{\rm rel},
{\rm{Cs}}}(\alpha) \approx +0.8\,.
\end{equation}
Therefore, the comparison of the clock transition in Hg$^+$
against a primary frequency standard tests a drift of \cite{Biz03}
\begin{equation}
\frac{\partial}{\partial t} \ln\frac{\nu_{\rm Cs}}{\nu_{\rm Hg}} =
\frac{\partial}{\partial t}\left( \ln\frac{\mu_{\rm Cs}}{\mu_{B}}+(2.0+0.8+3.2)\ln\alpha \right)\, , %= -0,24 \pm 7,2 \times 10^{-15} \textrm{ pro Jahr }
\end{equation}
whereas the $1S$-$2S$ experiment tests the drift of
\begin{equation}
\frac{\partial}{\partial t} \ln\frac{\nu_{\rm Cs}}{\nu_{\rm H}} =
\frac{\partial}{\partial t}\left( \ln\frac{\mu_{\rm Cs}}{\mu_{B}}+(2.0+0.8)\ln\alpha \right)\,. %= -0,48 \pm 1,2 \times 10^{-14} \textrm{ pro Jahr }
\end{equation}

For clarity, we set $x\equiv\frac{\partial}{\partial t}\ln\alpha$,
$y\equiv\frac{\partial}{\partial t}\ln\frac{\mu_{\rm
{Cs}}}{\mu_{B}}$ and get as experimental constraints
\begin{subequations}
\label{rels}
\begin{eqnarray}
\label{relsa}
y +6x & = & (0.2 \pm 7)
\times 10^{-15}\ \textrm{yr}^{-1} \ \ \textrm{ (Hg}^+\textrm{),} \\
\label{relsb} y+2.8x & = & (3.2 \pm 6.3) \times 10^{-15} \
\textrm{yr}^{-1} \ \ \textrm{ (H).}
\end{eqnarray}
\end{subequations}

\begin{figure}[t]
\begin{center}
\includegraphics[width=7.0cm]{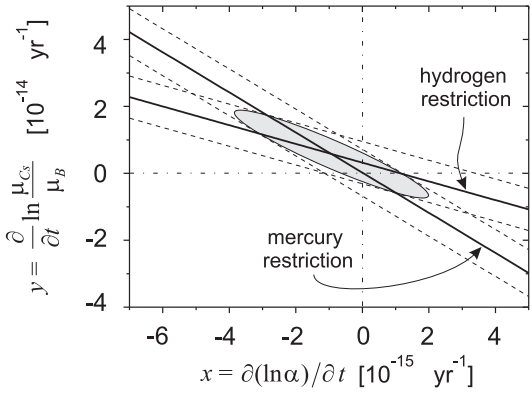}
\caption{\em Possible values for the relative drift rates $x$ and
$y$ and their experimental restrictions (dashed lines) as stated
by eqns. (7). The probability density for $x$ and $y$ to be the
true values is given by $P(\Delta x,\, \Delta y)\propto
\exp[-R(\Delta x,\, \Delta y)/2]$ with the distances from the
crossing point $\Delta x$ and $\Delta y$ and $R(\Delta x,\, \Delta
y)=(\Delta y+6\Delta x)^2/\sigma^2_{\rm Hg}+ (\Delta y+2.8\Delta
x)^2/\sigma^2_{\rm H}$. The resulting elliptical uncertainty
region, defined by $R(\Delta x,\, \Delta y)=1$, gives the standard
deviation when projected on either axis by integration over the
other. The uncertainties of the measured values are $\sigma_{\rm
H}=6.3\times10^{-15}$~yr$^{-1}$ and $\sigma_{\rm
Hg}=7\times10^{-15}$~yr$^{-1}$. } \label{bilddrifthgh}
\end{center}
\end{figure}

Solving for $x$ and $y $ yields the separate restrictions for the
drifts of $\alpha$ and $\mu_{\rm {Cs}}/\mu_B$ without any
assumptions of conceivable mutual correlations (see e.g.
\cite{Cal02}). In this sense, this evaluation is
model-independent. Figure \ref{bilddrifthgh} represents both
equations and their solution graphically.

For the relative drift of the fine structure constant at the end
of the second millennium, we deduce the limit of
\begin{equation}
x=\frac{\partial}{\partial t}{\ln \alpha} = (-0.9 \pm 2.9) \times
10^{-15} \ {\rm yr}^{-1}. \label{alph}
\end{equation}

The limit on the relative drift of $\mu_{\rm {Cs}}/\mu_{B}$ is
\begin{equation}
y=\frac{\partial}{\partial t} \ln \frac{\mu_{\rm {Cs}}}{\mu_{B}}=
(0.6 \pm 1.3) \times 10^{-14} \ {\rm yr}^{-1}. \label{magn}
\end{equation}

We deduce the uncertainties in expressions (\ref{alph}) and
(\ref{magn}) as projections of the ellipse
(Fig.~\ref{bilddrifthgh}) on the corresponding axes. This is
equivalent to performing  Gaussian propagation of uncertainties
when resolving Eqs.\,(7) for $x$ and $y$. Here, the measurement
results for Hg$^+$ and H are treated as uncorrelated even though
the drift rates $x$ and $y$ may be correlated \cite{Cal02}. The
given $1\sigma$-uncertainties for $x$ and $y$ in Eqs.~(\ref{alph})
and (\ref{magn}) incorporate both the statistical and systematic
uncertainties of the hydrogen and the mercury measurements. Both
limits (\ref{alph}) and (\ref{magn}) are consistent with zero. Meanwhile E. Peik and coworkers have added a precise drift measurement on a single trapped Yb ion to push the overall limit even further \cite{Peik}.

These results allow us to deduce a restriction for the relative
drift of the nuclear magnetic moments in $^{87}$Rb and $^{133}$Cs.
From 1998 to 2003, the drift of the ratio of the ground state
hyperfine frequencies in $^{87}$Rb and $^{133}$Cs has been
measured to be \cite{Mar03}

\begin{eqnarray}
\frac{\partial}{\partial t} \ln\frac{\nu_{\rm {Rb}}}{\nu_{\rm
{Cs}}}= (0.2 \pm 7.0) \times 10^{-16}\ \textrm{
yr}^{-1}.\label{clock}
\end{eqnarray}
\\

Substituting the corresponding dependencies $F_{\rm rel}(\alpha)$
for these transitions \cite{Mar03,Dzu99} yields
\begin{equation}
\frac{\partial}{\partial t} \ln\frac{\nu_{\rm {Rb}}}{\nu_{\rm
{Cs}}}=\frac{\partial}{\partial t}\left( \ln\frac{\mu_{\rm
{Rb}}}{\mu_{\rm {Cs}}}-0.53\ln\alpha\right). \label{magmom}
\end{equation}

Combining (\ref{alph}), (\ref{clock}), and (\ref{magmom}) we get
\begin{equation}
\frac{\partial}{\partial t} \ln\frac{\mu_{\rm {Rb}}}{\mu_{\rm
{Cs}}}=(-0.5\pm1.7)\times10^{-15}\ \textrm{ yr}^{-1},
\end{equation}
where the  same procedure as in Fig.\ref{bilddrifthgh} was used
with a diagram covering $x$ and $z\equiv\frac{\partial}{\partial
t} \ln(\mu_{\rm Rb}/\mu_{\rm Cs})$.

In conclusion, we have determined separate limits for the drift of
$\alpha$, $\mu_{\rm {Cs}}/\mu_{B}$ and $\mu_{\rm {Rb}}/\mu_{\rm
{Cs}}$ from laboratory experiments without assumptions of
conceivable correlations among them. All these limits are
consistent with zero. Quasar absorption spectra measured with the
Keck/HIRES spectrograph show a significant deviation between the
values of $\alpha$ today and 10 Gyr ago \cite{Mur03}. A
corresponding linear drift of $\alpha$ is smaller than the
uncertainty of our result, therefore it cannot be excluded.

We thank S.G.~Karshenboim  for fruitful discussions of this work.
N.K. acknowledges support from the AvH Stiftung. The work was
partly supported by DFG (grant No. 436RUS113/769/0-1) and RFBR.
The development of the FOM fountain was supported by Centre
National d'\'{e}tudes spatiales and Bureau National de
M\'{e}trologie.


\begin{thebibliography}{9.}
\bibitem[12]{remark} The result presented in
  \cite{Nie00a}  was inadvertently described as ``the
weighted mean value'' but was calculated without consideration of
day-averaged uncertainties.
\end{thebibliography}
\end{document}